\documentclass{jetpl}
\usepackage{graphicx}
\usepackage{amssymb}
\usepackage{amsmath}
\usepackage{color}
\usepackage{hyperref}
\twocolumn

\title{NMR shifts in $^3$He in aerogel induced by demagnetizing fields}
\rtitle{NMR shifts in $^3$He in aerogel induced by demagnetizing fields}
\sodtitle{NMR shifts in $^3$He in aerogel induced by demagnetizing fields}

\author{V.\,V.\,Dmitriev$^+$\thanks{e-mail: dmitriev@kapitza.ras.ru},\,M.\,S.\,Kutuzov$^*$,\,A.\,A.\,Soldatov$^{+,\times}$,\,A.\,N.\,Yudin$^{+,\circ}$}
\rauthor{V.\,V.\,Dmitriev,\,M.\,S.\,Kutuzov,\,A.\,A.\,Soldatov,\,A.\,N.\,Yudin}
\sodauthor{V.\,V.\,Dmitriev,\,M.\,S.\,Kutuzov,\,A.\,A.\,Soldatov,\,A.\,N.\,Yudin}

\address{$^+$P.\,L. Kapitza Institute for Physical Problems of RAS, 119334 Moscow, Russia}
\address{$^*$Metallurg Engineering Ltd., 11415 Tallinn, Estonia}
\address{$^\times$Moscow Institute of Physics and Technology, 141700 Dolgoprudny, Russia}
\address{$^\circ$National Research University Higher School of Economics, 101000 Moscow, Russia}

\dates{\today}{*}

\begin{document}

\abstract{Magnetic materials generate demagnetizing field that depends on geometry of the sample and results in a shift of magnetic resonance frequency. This phenomenon should occur in porous nanostructures as well, e.g., in globally anisotropic aerogels. Here we report results of nuclear magnetic resonance (NMR) experiments with liquid $^3$He confined in anisotropic aerogels with different types of anisotropy (nematic and planar aerogels). Strands of aerogels in pure $^3$He are covered by a few atomic layers of paramagnetic solid $^3$He which magnetization follows the Curie-Weiss law. We have found that in our samples the NMR shift in solid $^3$He is clearly seen at ultralow temperatures and depends on value and orientation of the magnetic field. The obtained results are well described by a model of a system of non-interacting paramagnetic cylinders. The shift is proportional to the magnetization of solid $^3$He and may complicate NMR experiments with superfluid $^3$He in aerogel.}

\maketitle

\section{Introduction}As it was shown by C.~Kittel \cite{kit48}, demagnetizing fields may result in an additional frequency shift in magnetic resonance experiments at large values of the sample magnetization. The spin susceptibility of liquid $^3$He is small, and in this case Kittel shifts in bulk samples were observed in experiments with spin polarized $^3$He \cite{Tast, Cand} or with thin $^3$He films \cite{Bozler, free88}. In the latter case the shift is due to the presence of few ($\sim2$) atomic layers of solid paramagnetic $^3$He adsorbed on the surface. In result, the overall nuclear magnetic resonance (NMR) signal from $^3$He, containing both liquid and solid components, is observed as a single NMR line (due to fast spin exchange mechanism \cite{free88}) with the frequency shift as weighted average of those in liquid and solid fractions of $^3$He.
Solid layers follow the Curie-Weiss law \cite{aho78,saul05,coll09} and their magnetization (as well as the Kittel shift) at low temperatures may be large. The Kittel shift also may be observable in normal $^3$He confined in different nanostructures, e.g., in aerogels consisting of nanostrands. In globally isotropic aerogel the average shift is zero, but it may appear in the anisotropic sample.

Here we study the Kittel effect in pure liquid $^3$He in two different globally anisotropic nanostructures called below nematic and planar aerogels.  Nematic aerogel consists of nearly parallel strands, while in planar aerogel the strands are uniformly distributed in a plane perpendicular to the symmetry axis. The solid $^3$He adsorbed on the strands can be considered as a system of independent cylindrical surfaces oriented either along one direction (in nematic aerogel) or chaotically distributed in the plane (in planar aerogel). We use this model to calculate the Kittel frequency shifts in $^3$He for both cases and to interpret our experimental results.

\section{Theoretical model}
The total magnetic susceptibility of $^3$He in aerogel is a sum of the susceptibilities of liquid ($\chi_l$) and solid ($\chi_s$) $^3$He:
\begin{equation}\label{susc}
\chi=\chi_l+\chi_s=\chi_l+\frac{C_s}{T-\Theta},
\end{equation}
where $C_s$ is the Curie constant, $\Theta$ is the Curie temperature of solid $^3$He, $\chi_i\equiv M_i^{aero}/H$, $\bf H$ is an external magnetic field, $M_i^{aero}$ is a total magnetic moment of liquid ($i=l$) or solid ($i=s$) $^3$He per unit volume of the \textit{aerogel} sample. $\chi_l$ is temperature independent in normal liquid $^3$He and may only decrease with temperature in superfluid $^3$He \cite{VW}, so at low temperatures the NMR signal from solid $^3$He can prevail over that from liquid $^3$He.

In experiments with pure $^3$He in aerogel the common NMR line has the following frequency shift \cite{free88}:
\begin{equation}\label{fastexch}
\Delta\omega^\prime=\frac{\chi_l\Delta\omega_l+\chi_s\Delta\omega_s}{\chi_l+\chi_s},
\end{equation}
where $\Delta\omega_l$ is a frequency shift in liquid $^3$He, $\Delta\omega_s$ is a Kittel frequency shift in solid $^3$He. Here all the shifts are measured from the Larmor frequency $\omega_L=\gamma H$, where $\gamma$ is the gyromagnetic ratio of $^3$He.

In the first approximation the solid $^3$He adsorbed on the strands of nematic and planar aerogels is a combination of cylindrical surfaces. The frequency shift in solid $^3$He at a separate strand \cite{Cand} is
\begin{equation}\label{kitstr}
\Delta\omega_s=\pi\gamma M_s^{cyl}\left(2-3\sin^2\varphi\right),
\end{equation}
where $M_s^{cyl}$ is a magnetization of the solid $^3$He on a cylinder surface, $\varphi$ is an angle between $\bf H$ and the cylinder axis. In nematic aerogel strands are almost parallel to one another, so the mean frequency shift in $^3$He adsorbed on nematic aerogel is given by
\begin{equation}\label{kitnem}
\Delta\omega_s=\pi\gamma\frac{\chi_s}{s_V\delta}H\left(2-3\sin^2\varphi\right),
\end{equation}
where $s_V$ is an effective surface area per unit volume of the aerogel, $\delta$ is a thickness of solid $^3$He layers. The shift is positive for $\varphi=0$, negative for $\varphi=\pi/2$, while the ratio of the corresponding absolute values is 2.

In planar aerogel strands are parallel to the distinguished plane. After averaging over angular distribution of the non-interacting strands in the plane we get the following value of the frequency shift:
\begin{equation}\label{kitpl}
\Delta\omega_s=\pi\gamma\frac{\chi_s}{s_V\delta}H\left(\frac{3}{2}\sin^2\varphi-1\right),
\end{equation}
where $\varphi$ is an angle between $\bf H$ and the normal to the plane. In contrast to the case of nematic aerogel the shift is negative for $\varphi=0$, positive for $\varphi=\pi/2$, while the ratio of the corresponding absolute values is still 2.

\section{Samples and methods}
In the experiments as a nematic aerogel we have used a nanomaterial called nafen \cite{anf} produced by ANF Technology Ltd. It consists of Al$_2$O$_3$ strands which are oriented along the same direction, have diameters $d\approx9$\,nm \cite{asad15} and length $\sim1$\,cm. We had two samples of nafen: nafen-243 with overall density $\rho=243$\,mg/cm$^3$, porosity $p=93.9$\%, characteristic separation of strands $\ell\approx40$\,nm and nafen-910 with $\rho=910$\,mg/cm$^3$, $p=78$\%, $\ell\approx20$\,nm. The sample of nafen-910 was obtained from nafen with density of 72\,mg/cm$^3$ by a technique  described in Ref.~\cite{vol17}.

The sample of planar aerogel was produced from an aluminum silicate (mullite) nematic aerogel consisting of strands with $d\approx10$\,nm (see Ref.~\cite{dmit18}). It is a fibrous network in the plane with $p=88$\%, $\rho=350$\,mg/cm$^3$, and with characteristic lengths of separate strands of $\sim1$\,$\mu$m which is much bigger than their diameters.

The spin diffusion measurements in normal $^3$He confined by these nanostructures \cite{dmit18,dmit15} confirm their strong anisotropy.

Samples of nafen had a form of cuboid with a side 4\,mm, the sample of planar aerogel was a stack of three plates with thickness $\approx1$\,mm and sizes $4\times4$\,mm. They were placed in the separate cells of our experimental chamber with a filling factor $\sim85$\%. The experimental chamber was made from Stycast-1266 epoxy resin and was similar to that described in Ref.~\cite{ask12}.

\begin{figure}[t]
\centerline{\includegraphics[width=\columnwidth]{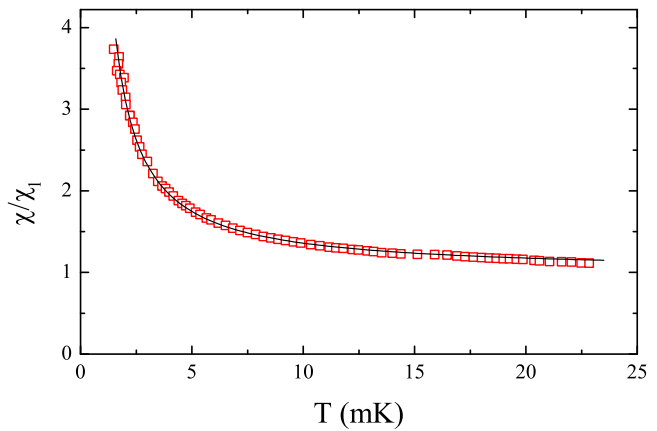}}
\caption{Fig.\,\thefigure. 
The total magnetic susceptibility of pure $^3$He in nafen-243 normalized to $\chi_l$ in normal $^3$He. $\varphi=0$, $P=7.1$\,bar, $\omega_L/(2\pi)=880$\,kHz. The solid curve is a fit to Eq.~\eqref{susc}. $\Theta=0.37$\,mK, $\chi_s/\chi_l\approx2.7$ at $T=T_c=1.643$\,mK}
\label{susc_naf243}
\end{figure}
\begin{figure}[t]
\centerline{\includegraphics[width=\columnwidth]{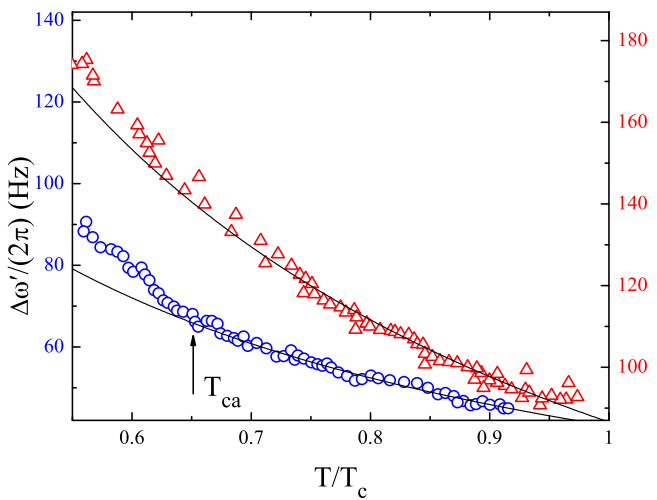}}
\caption{Fig.\,\thefigure. 
The original frequency shifts of pure $^3$He in nafen-243 in lower ($\omega_L/(2\pi)=361$\,kHz, circles, left $y$-axis) and higher ($\omega_L/(2\pi)=880$\,kHz, triangles, right $y$-axis) magnetic fields. $\varphi=0$, $P=7.1$\,bar. The arrow indicates the superfluid transition temperature of $^3$He in aerogel $T_{ca}\approx0.65T_c$. Solid curves are fits to Curie-Weiss law at $T>T_{ca}$}
\label{df_naf243}
\end{figure}

Experiments were carried out using linear continuous wave NMR in magnetic fields 24--402\,Oe (corresponding NMR frequencies are 78--1303\,kHz) at pressures s.v.p.--29.3\,bar. The purity of used $^3$He was about 0.01\% in experiments with nafen and 0.07\% in experiments with planar aerogel, which for the surfaces inside the experimental chamber dominated by the heat exchanger with area of $\approx40$\,m$^2$ corresponds to preplating of aerogel strands with $\sim0.1$ and $\sim0.6$ atomic layers of $^4$He respectively. We were able to rotate $\bf H$ by any angle $\varphi$ defined in the previous section. Additional gradient coils were used to compensate the magnetic field inhomogeneity. The necessary temperatures were obtained by a nuclear demagnetization cryostat and measured by a quartz tuning fork. Below the superfluid transition temperature ($T_c$) of bulk $^3$He  the fork was calibrated by Leggett frequency measurements in bulk $^3$He-B. Above $T_c$ the temperature was determined in assumption that the resonance linewidth of the fork in normal $^3$He is inversely proportional to the temperature \cite{fork}.

\section{Results and discussions}
In all samples the measured magnetic susceptibility, determined from the intensity of the NMR absorption line, has a clear Curie-Weiss behavior (Fig.~\ref{susc_naf243}). Due to the fast exchange between liquid and solid $^3$He atoms we observe a single NMR line. In Fig.~\ref{df_naf243} examples of temperature dependencies of the frequency shift in $^3$He in nafen-243 are shown. In high magnetic field the shift (triangles) is mostly determined by the Kittel shift from the surface solid $^3$He, while at lower field the kink is observed on the data (circles) indicating a transition to superfluid $^3$He. At $T<T_c$ the magnetic susceptibility of solid $^3$He in the sample $\chi_s\gg\chi_l$, so when liquid $^3$He in aerogel is normal ($\Delta\omega_l=0$) from Eqs.~(\ref{susc},\ref{fastexch},\ref{kitnem}) it follows that $\Delta\omega^\prime\approx\Delta\omega_s\propto1/(T-\Theta)$ (solid lines in Fig.~\ref{df_naf243}). In superfluid $^3$He the frequency shift is usually inversely proportional to the magnetic field $\Delta\omega_l\propto1/H$ \cite{VW}, while in the solid $^3$He on the aerogel strands $\Delta\omega_s\propto H$ according to Eqs.~(\ref{kitnem},\ref{kitpl}). Therefore, low magnetic fields in NMR experiments allow to get rid of the Kittel effect originating from anisotropy of the aerogel and to investigate purely superfluid properties of $^3$He in aerogel. On the other hand, in high magnetic fields the superfluid frequency shift can be significantly suppressed with respect to the Kittel shift.

\begin{figure}[t]
\centerline{\includegraphics[width=\columnwidth]{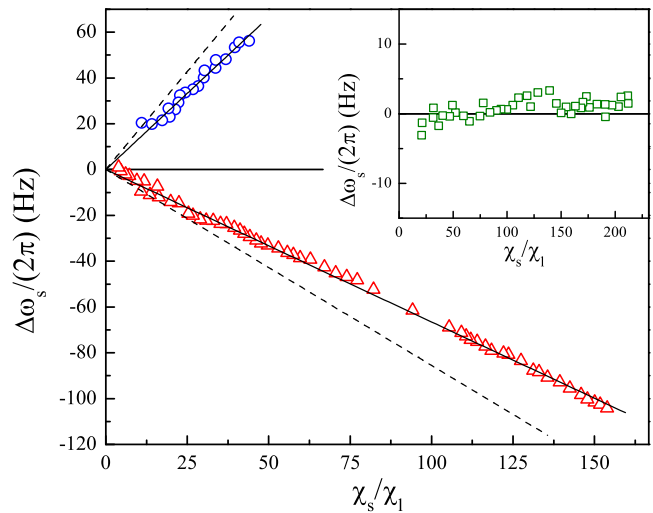}}
\caption{Fig.\,\thefigure. 
The Kittel shift in solid $^3$He in nafen-910 versus $\chi_s/\chi_l$ at $P=7.1$\,bar for $\varphi=0$ (circles) and $\varphi=\pi/2$ (triangles) (at $P=29.3$\,bar for $\varphi\approx55^\circ$ in the inset). $\omega_L/(2\pi)=361.4$\,kHz ($\omega_L/(2\pi)=78.4$\,kHz in the inset). The dependencies are implicit functions of $T$, e.g., triangles correspond to temperatures from 23\,mK down to $\sim0.6$\,mK. Solid lines are linear fits with the ratio of slopes $\approx1.9$. Dashed lines are theoretical predictions (see text)}
\label{df_naf910}
\end{figure}
\begin{figure}[t]
\centerline{\includegraphics[width=\columnwidth]{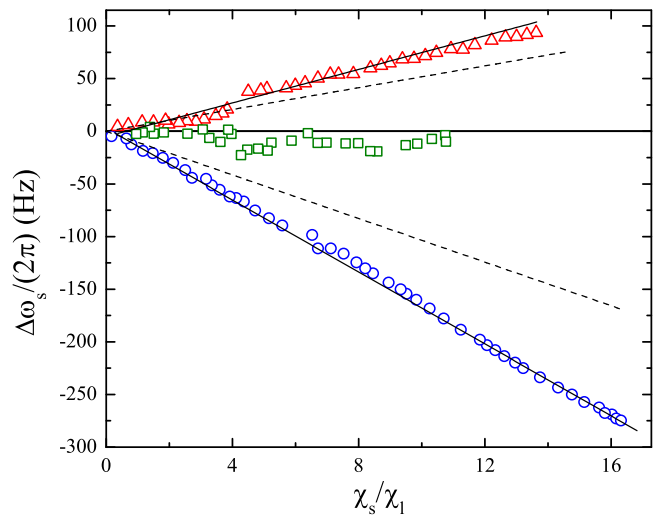}}
\caption{Fig.\,\thefigure. 
The Kittel shift in solid $^3$He in planar aerogel versus $\chi_s/\chi_l$ at s.v.p. ($P\approx0$\,bar) for $\varphi=0$ (circles), $\varphi=\pi/2$ (triangles), and $\varphi\approx55^\circ$ (squares). Triangles and squares ($\omega_L/(2\pi)=588.6$\,kHz) are recalculated to match the Larmor frequency of circles ($\omega_L/(2\pi)=1303.2$\,kHz). The dependencies are implicit functions of $T$, e.g., circles correspond to temperatures from 14\,mK down to $\sim0.5$\,mK. Solid lines are linear fits with the ratio of slopes $\approx2.1$. Dashed lines are theoretical predictions (see text)}
\label{df_planar}
\end{figure}

The Kittel effect is more clearly manifested in nafen-910 which is denser than nafen-243. Superfluidity of $^3$He in presence of solid $^3$He on the aerogel strands is completely suppressed in nafen-910 \cite{mag}, so using Eq.~\eqref{fastexch} we can determine $\Delta\omega_s$ from measurements of $\Delta\omega^\prime$ down to the lowest attained temperatures (see Fig.~\ref{df_naf910}). The shift is positive in the magnetic field parallel to the nafen anisotropy axis ($\varphi=0$) and negative in the transverse direction of the field ($\varphi=\pi/2$). The absolute value of the ratio of the corresponding shifts is $\approx1.9$ which is in a good agreement with Eq.~\eqref{kitnem}, while the data in the inset (squares) demonstrate the absence of the shift at $\sin^2\varphi=2/3$.

To estimate the expected value of $\Delta\omega_s$ using Eq.~\eqref{kitnem}, we need to know values of $\chi_s$, $s_V$, and $\delta$.
First, $\chi_s$ can be found from measurements of the total magnetization of the sample that allows to determine $\chi_s/\chi_l$. We note that $\chi_l=p\chi_l^{bulk}\approx4.25\cdot10^{-8}$\,emu, where $p=0.78$ is the porosity of nafen-910 and $\chi_l^{bulk}=5.45\cdot10^{-8}$\,emu is the magnetic susceptibility in bulk normal $^3$He at $P=7.1$\,bar. Second, $s_V$ can be estimated in the assumption that the nafen strands are ideal cylinders. In this case $s_V=\frac{4}{d}\frac{\rho}{\rho_0}\approx102$\,m$^2$/cm$^3$, where $d\approx9$\,nm is a strand diameter, $\rho=910$\,mg/cm$^3$ is nafen-910 density, and $\rho_0=3.95$\,g/cm$^3$ is Al$_2$O$_3$ density. Third, the solid $^3$He on the surfaces at $P=7.1$\,bar has a coverage of $\approx1.6$ atomic layers \cite{sch87}. The thickness of one layer may be estimated as $3.5$\,\r{A} (the lattice period of $^3$He crystal at low temperatures) that gives $\delta\approx5.5$\,\r{A}. In result, theoretical predictions according to Eq.~\eqref{kitnem} are plotted in Fig.~\ref{df_naf910} as dashed lines which are surprisingly close to the experimental results, considering a rather bad accuracy in estimations of $\chi_s$, $s_V$, and $\delta$ ($\pm20$\%).

The frequency shift measurements in solid $^3$He in planar aerogel (see Fig.~\ref{df_planar}) are also in agreement with the theory (Eq.~\eqref{kitpl}). The shift for $\varphi=0$ is negative and $\approx2.1$ larger than that for the field along the plane ($\varphi=\pi/2$) which is positive. At $\sin^2\varphi=2/3$ the shift is zero as expected from Eq.~\eqref{kitpl}. Estimation of the shift values in planar aerogel using Eq.~\eqref{kitpl} gives the same order of magnitude as in experiments. In this case $\chi_l=p\chi_l^{bulk}\approx3.11\cdot10^{-8}$\,emu (where $p=0.88$ and $\chi_l^{bulk}=3.53\cdot10^{-8}$\,emu at s.v.p.), $s_V=\frac{4}{d}\frac{\rho}{\rho_0}\approx47$\,m$^2$/cm$^3$ (where $d\approx10$\,nm, $\rho=350$\,mg/cm$^3$ is planar aerogel density, and $\rho_0\approx3$\,g/cm$^3$ is mullite density), and $\delta\approx2.8$\,\r{A} at s.v.p. for 0.8 atomic layers of solid $^3$He \cite{sch87} (here we assume that $\sim0.6$ atomic layers of solid $^3$He is replaced by $^4$He \cite{free90} due to a rather ``dirty'' $^3$He with 0.07\% of $^4$He used in the experiment).

\section{Conclusions}
We have observed NMR shifts due to the Kittel effect in $^3$He confined in aerogel-like nanostructures with different types of the global anisotropy and demonstrated that values of the shift well agree with the theoretical expectations. At ultralow temperatures even in moderate magnetic fields these shifts may be large enough to mask the $^3$He superfluid transition but can be avoided by using lower magnetic fields or by choosing the proper angle between the axis of the anisotropy and the magnetic field.

This work was supported by grant of the Russian Science Foundation (project \#\,18-12-00384).

\end{document}